# Design of a High Step-up DC-DC Power Converter with Voltage Multiplier Cells and Reduced Losses on Semiconductors for Photovoltaic Systems

Mamdouh L. Alghaythi
Department of Electrical Engineering and Computer Science
University of Missouri
Columbia, USA
mlagzd@mail.missouri.edu

Robert M. O'Connell
Department of Electrical Engineering and Computer Science
University of Missouri
Columbia, USA
OConnellR@missouri.edu

Naz E. Islam
Department of Electrical Engineering and Computer Science
University of Missouri
Columbia, USA
islamn@missouri.edu

*Abstract*—A high step-up dc–dc converter based on an isolated dc-dc converter with voltage multiplier cells for photovoltaic systems is essentially introduced in this paper. The proposed converter can provide a high step-up voltage gain. The switch voltage stress and losses on semiconductors are significantly reduced through this work. Furthermore, the proposed converter can reliably offer and provide continuous input current which can be basically used for integrating photovoltaic systems to convert 30 V to 480 V dc bus. The ripple on the input current is minimized due to the isolated converter, and the proposed converter is fed by a single input voltage. The operation modes and the characteristics of the aforementioned converter are thoroughly analyzed. The components selection, simulation results and experiment results are mainly verified by using MATALB Simulink. Consequently, a 360 W hardware prototype is implemented to validate the design and the theory.

## I. Introduction

The growing demand in the recent years for renewable energy systems, photovoltaic systems, fuel cells, and semiconductor industry requires wide use of high voltage gain dc-dc power electronic converters. In the past years, the high intensity discharge lamp ballast in automotive headlamps used high voltage gain dc-dc [1]. In the recent time, photovoltaic (PV) cells sources are becoming more extensive with the growing energy demand. The dc voltage of a single photovoltaic panel is essentially low and around 50 V. In other words, the output voltage of any renewable energy system, such as a single PV, is not suitable to step up a high output voltage and output power. Furthermore, the conventional boost, buck-boost, and flyback converters must fundamentally have high large duty ratios to increase the output voltage and get a high voltage gain [2]. However, by increasing the high cycles ratios, the ripples on the input current, and the stresses on the voltage switches are going to consequently increase [7]-[10]. Fig. 1 shows that a typical photovoltaic system that a PV source, high step up DC-DC converters are used to increase the low voltage of a PV source from 30 V to 480 V which essentially would be converted to an AC voltage through an inverter to the load.

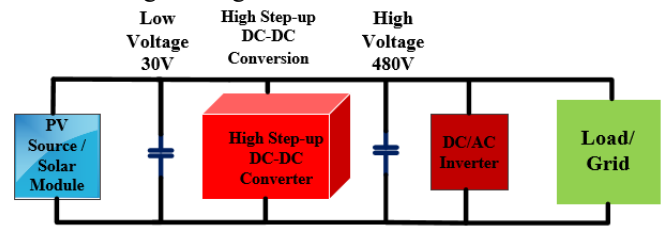

Fig. 1. A typical high step-up dc-dc converter with a photovoltaic system

Large ripples on the input currents, large ripples on the output voltage and the voltage stresses on the switches would inevitably lower the efficiency and inescapably the productivity of the converter [3]. In addition, since the diode does not conduct for a long time, a diode reverse recovery might essentially cause some issues. One way to avoid and reduce the large ripples on the current, and diode reverse recovery is to use a coupled inductor and a transformer [12]. Even though cascading and adding additional non-isolated dc-dc converters would increase the output voltage, it may eventually lower the efficiency. Furthermore, another disadvantage of the cascaded conventional converter is that the output diode may basically experience a high voltage stress on the switches [15]-[22]. Moreover, to accomplish a high voltage gain in the isolated converters, an increase in the turns ratio is required. For if not, the number of winding turns in the transformers may cause the leakage inductance, and therefore may cause the voltage spikes [4]. Despite using the switched capacitors, which would decrease the voltage stresses on the switches, it may cause losses in the capacitor voltages [5]. One way to overcome these problems is to step up the output voltage, minimize the voltage stresses on the switches and minimize the duty ratio that is essentially using voltage multiplier cells (VMCs) in the conventional converter [6]. Thus, since using voltage multiplier cells would experience discontinuous input current, new topologies are introduced here to overcome these problems [11]. Moreover, because of the leakage inductance, and limited resistance components, conventional boost and flyback converters are not suitable to attain a high voltage gain with a high efficiency [13,14]. The proposed converter has a winding transformer, coupled inductors, two switches, a single input voltage and a voltage multiplier cell. The paper will provide the analysis of the operation modes in part A of section II, mathematical expression in part B of section II , simulation results and components selection in part C of section II experiment results, the losses and the efficiency of the proposed converter in section IV and conclusion in section V.

## II. ANALYSIS OF THE PROPOSED CONVERTER

The proposed converter is fed by a single source as shown in figure 2. The voltage multiplier cells play a remarkable role in increasing the voltage gain. It can be observed that the voltage transformation ratio is essentially based on switch duty ratios and the number of voltage multiplier cells. The converter is fundamentally operating in continuous conduction mode and the components are principally ideal. Moreover, it can be observed that the proposed converter has two stages, two switches, a winding transformer, inductors, diodes, and capacitors. The active switches are simultaneously controlled by one control signal as shown in fig. 3. As a result, the proposed converter has three modes of operation.

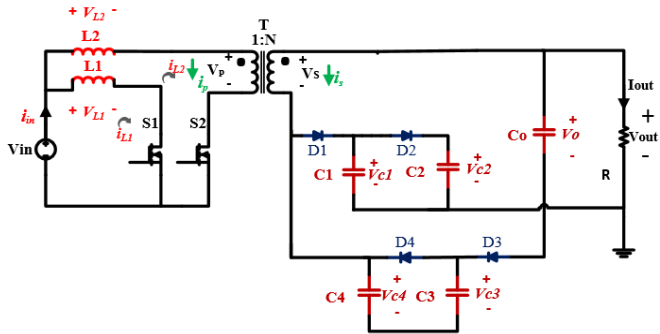

Fig. 2. The proposed high step up dc-dc converter

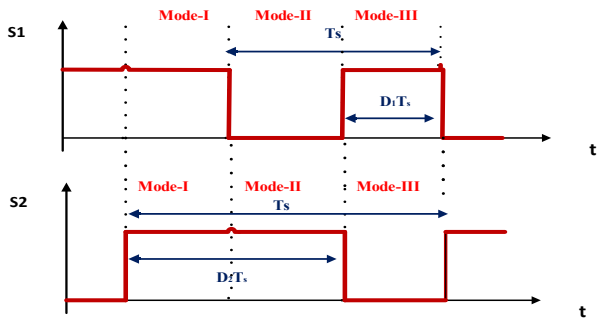

Fig. 3. The switching signals for the proposed converter

*A. Investigation and Analysis of the Modes of Operation*

The three modes of operation are essentially analyzed and studied in this section.

*1) Mode-I*

In this mode, as shown in fig. 4 switches S1 and S2 are together ON. Moreover, all diodes are reverse biased, and they are OFF. Inductors, and the winding transformer are charging by the single source. As it is in the figure, the load has been supplied by the output capacitor.

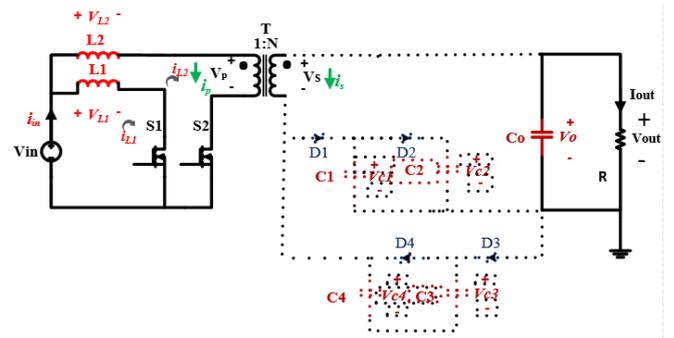

Fig. 4. Mode I of the proposed high step up dc-dc converter

*2) Mode-II*

For this mode, S1 is OFF, and S2 is ON as shown in fig. 5. D2 and D4 are reverse biased, and they are OFF. But D1 and D3 are forward biased, and they are ON. Thus, the inductor current for L1 plays a remarkable role in charging C1 and C3. However, the other current is essentially going to discharge C2 and C4.

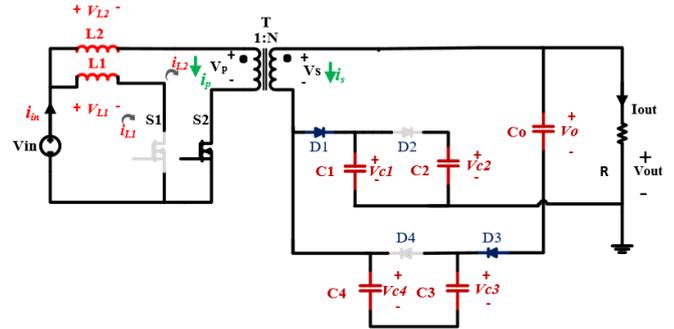

Fig. 5. Mode II of the proposed high step up dc-dc converter

*3) Mode-III*

For this mode, S1 is ON, and S2 is OFF as shown in fig. 6. D2, D4 are forward biased, and they are ON. But D1 and D3 are reverse biased, and they are OFF. Hence, D2 and D4 conduct, but D1 and D3 do not obviously conduct. The inductor current for L2 charges C2 and C4 but the inductor current for L1 discharges C1 and C3.

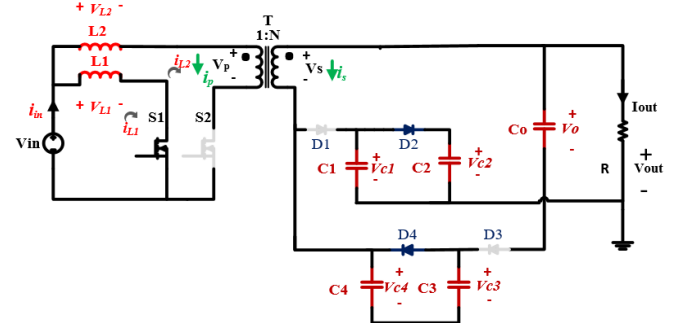

Fig. 6. Mode III of the proposed high step up dc-dc converter

Thus, there are some considerations in analyzing the operation modes; the components are ideal, the system is in the steady state, inductor currents are in continuous conduction mode and they categorically do not reach zero.

Table I. Simulation and components selection

| Parameter | Value |
|---|---|
| Load Resistance | 640 Ω |
| $L_1$ and $L_2$ | 120 µH |
| (C1,C2,C3 and C4) | 20 µF |
| Duty cycle of switches $S_1$ and $S_2$ (D) | 75% or 0.75 |
| Switching Frequency - $f_{sw}$ | 100 kHz |
| Input Voltage | 30 V |
| Output Voltage | 480 V |

Table II. Experiment and components ratings

| Component | Name | Rating | Part No |
|---|---|---|---|
| Capacitors | $C_1$, $C_2$, $C_3$ and $C_4$, | 20 µF | TVA1966-E3 |
| MOSFET | $S_1$ and $S_2$ | 150 V, 43 A, Rds(on)=7.5 mΩ, TO220FP-3 OptiMOS 3 | IPA075N15N3GXKSA1 |
| Diode | $D_1$, $D_2$, $D_3$ and $D_4$ | $I_f$=40A, $V_f$=0.61 V $V_R$= 80-100 V | 40CPQ100 |

The inductor current is 6A and can be respectively written as

$$i_{L(avg)} = \frac{2 \times i_{out}}{(1-D)} \quad (7)$$

The equations above have been analyzed and calculated to approve the simulation model and validate the experiment results.

### B. Analysis of the Voltage Gain and the Equations

First, by charging and discharging the voltage multiplier cell capacitors, the input power is transferred to the output side. The inductors voltages are calculated as

$$V_{in} = \frac{di_{L1}}{dt} \quad (1)$$

$$V_{in} = \frac{di_{L2}}{dt} \quad (2)$$

$$V_{in} = V_L \quad (3)$$

It can be observed the voltage across capacitors can be written as

$$V_{c2} = V_{c3} = V_{in} \times \frac{1}{(1-D)} \quad (4)$$

$$V_{c1} = V_{c4} = V_{in} \times \frac{2}{(1-D)} \quad (5)$$

Hence, the output voltage can be written as

$$V_{out} = V_{in} \times \frac{4}{(1-D)} \quad (6)$$

### C. Simulation Results and Components Slection

The simulation model of the proposed converter has been designed by using MATLAB Simulink. In addition, the selection of the parameters are given in Table I. The input voltage is essentially 30 V, load resistance is 640 Ω, the inductors are 120 µH, the voltage multiplier capacitors are 20 µF, the switching frequency is 100 kHz, and the duty ratio is 0.75.

The simulation has been done to meet the specifications and to boost the proposed converter to a higher voltage gain. Furthermore, the inductor currents are basically equal. Hence, the average inductor currents, and the voltage across inductor are explored through the simulation and the equations. From fig. 7, the inductor voltage is 30 V, and the input current ripple is successfully decreased. The input current is around 12.5A, and the input current ripple is 1.8A.

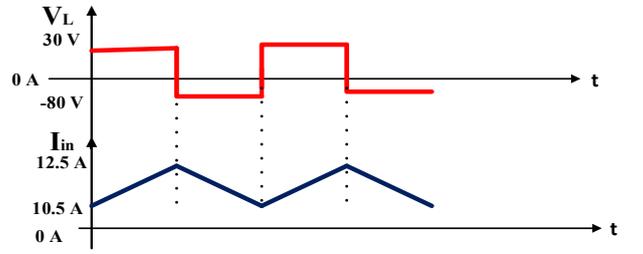

Fig. 7. The inductor voltage $V_L$[V], and the input current $I_{in}$[A] vs. time [S]

$$V_{sw} = V_{in} \times \frac{1}{(1-D)} \quad (8)$$

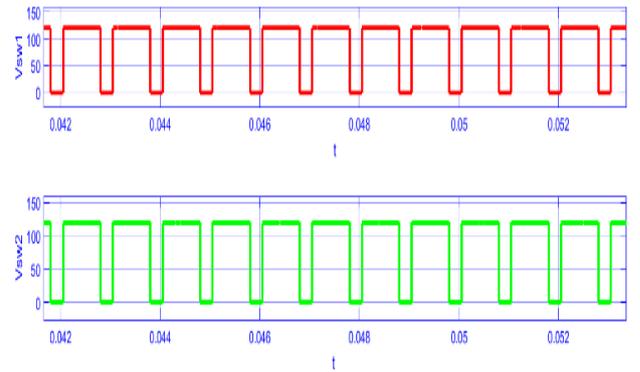

Fig. 8. The switch voltages, $V_{sw1}$ [V] and $V_{sw2}$[V] vs. time [S]

### III. EXPERIMENT RESULTS

A hardware prototype of the proposed converter was designed to validate the proposed converter operation. In addition, the specifications of the components used for building hardware prototype are specified as shown in Table II above. For the switches, it can be observed from fig. 8 above that the voltage stress across switches has been reduced to 120 V. Thus, the amount of the reduced switch voltage to the output voltage of the proposed converter is typically small. The voltage stress on the active switches can be calculated as

$$V_{sw} = V_{in} \times \frac{1}{(1-D)} = 120V \qquad (9)$$

The main reason of having a higher diode voltage is logically that the diode depends on the voltage multiplier capacitors. Moreover, the diode currents are completely equal to the output current. Moreover, the diode voltage is twice the switch voltage Thus, the voltage stress across the diodes has been minimized to 240 V as in fig. 9 and can be calculated as

$$V_D = V_{in} \times \frac{2}{(1-D)} \qquad (10)$$

The average diode current is the same as the output current and essentially equal to 0.75 A. By going back to the operation modes, it can be explored that the diodes only conduct in the second and the third modes which result in an inequal diode voltages. The only diode that is always conducting is the output diode which is here D4.

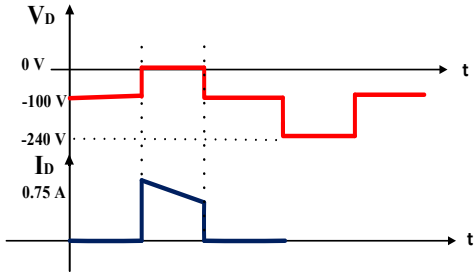

Fig. 9. The diode voltage $V_D$[V] and the diode current $I_D$[A] vs. time [S]

The output has been successfully increased to 480 V as in fig. 10, and it is previously calculated as

$$V_{out} = V_{in} \times \frac{4}{(1-D)} = 30 \times \frac{4}{(1-0.75)} = 480\ V \qquad (11)$$

It can be observed from fig. 10, the output voltage reaches the steady state at time of 0.3 S.

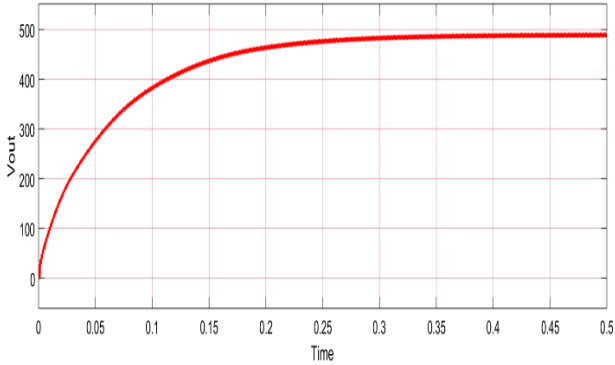

Fig. 10. The 0utput voltage $V_{out}$ [V] vs. time [S]

## IV. THE LOSSES AND THE EFFICIENCY OF THE PROPOSED CONVERTER

The conduction losses of inductor can be calculated by using copper wire DC loss. The wire loss caused by DC resistance as

$$P_{DCR}(W) = I_{rms}^2 \times DCR \qquad (12)$$

Where $I_{rms}$ is the rms value of the peak current applied to the inductor L1 and L2, and DCR is DC resistance of the inductor

$$P_{sw\,cond} = (R_{DS,S1} \times I_{S1,rms}^2) + (R_{DS,S2} \times I_{S2,rms}^2) \qquad (13)$$

The switching losses for both switches can written as

$$P_{sw\,loss} = \left[\frac{V_{S1} \times I_{L1,avg} \times (T_{on,s1} + T_{off,s1}) \times f_{SW}}{2}\right] + \left[\frac{V_{S2} \times I_{L2,avg} \times (T_{on,s2} + T_{off,s2}) \times f_{SW}}{2}\right] \qquad (14)$$

Fig. 11 shows the losses of the selected components, and it can be noticed that the remarkable losses happen in the diodes which are around 53% of the losses, 22% of the losses are in the inductors, 20% of the losses are in the MOSFETs, and 5% of the losses are in the capacitors.

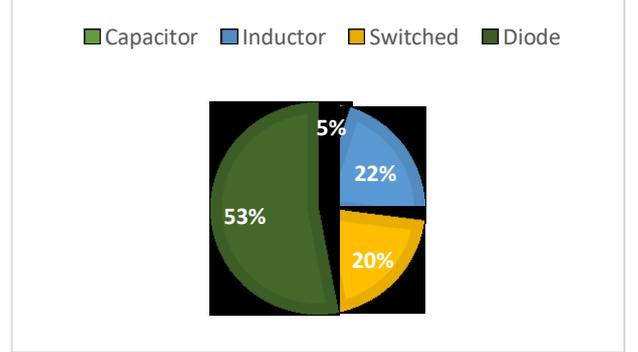

Fig. 11. The losses of the components in a percentage form

The maximum efficiency is essentially high and is 96 % as shown in fig. 12, and it can be also improved by selecting efficient components. Moreover, it can be noticed that the highest efficiency is accomplished at a power of 173 W. Therefore, by increasing the input power, the efficiency will slightly decrease.

$$\eta = \frac{P_{out}}{P_{in}} = \frac{V_{out} \times I_{out}}{V_{in} \times I_{in}} \times 100\% = 96\ \% \qquad (15)$$

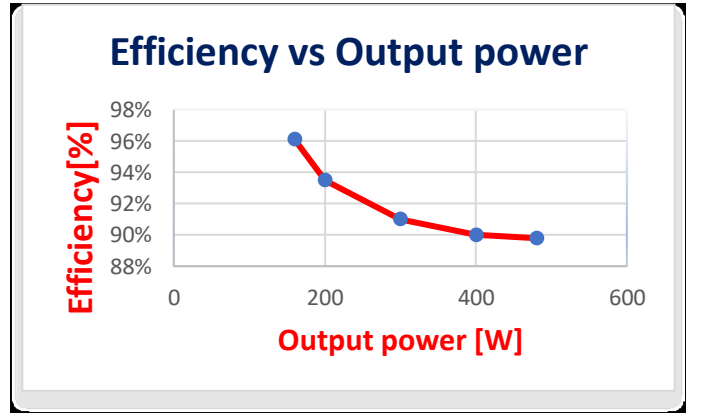

Fig. 12. Efficiency of the proposed converter vs. the output power

Fig. 13 shows the experimental waveforms of the switch voltages and the input current. Moreover, the switch voltages are both the same and are nearly 120V. The input current is approximately 12.5 A. it can be observed that the ripple on the input current has been reduced as it was previously calculated.

Fig. 14 shows that the experimental waveforms of the inductor current which is approximately 6 A. Therefore, the experimental waveforms validate the equations and the simulation where the stress voltage across the components are essentially minimizing. Hence, it can be noticed that the inductor currents are 180 degrees out of the phase.

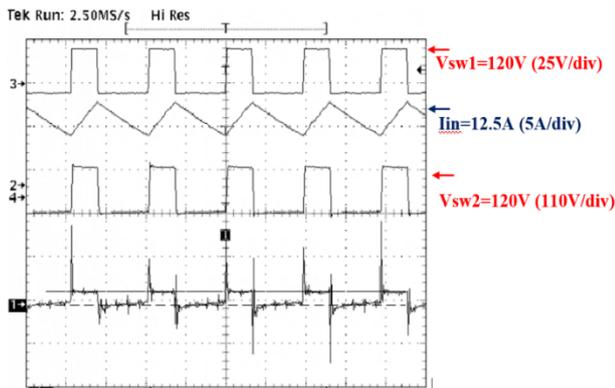

Fig. 13. The experiment waveforms of the switch voltages and the input current

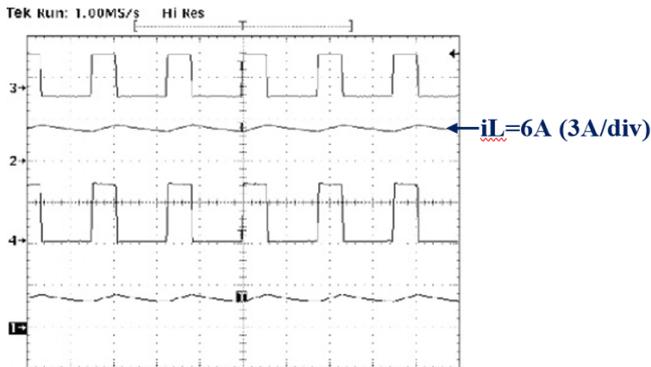

Fig. 14. The experiment waveform of the inductor current

## V. CONCLUSION

A high step up DC-DC converter using voltage multiplier cells was designed and analyzed. The proposed converter was successfully demonstrated to boost 30V to 480V. Moreover, the aforementioned converter reliably offered a high voltage gain with a reduced duty ratio. The voltage stresses across semiconductors were greatly reduced. The ripples on the input current were minimized and mitigated, and the losses on the leakage inductance were reduced. The operation modes, and the equations have been analyzed and explained. The simulation model and the hardware prototype were introduced to validate this research. The achieved maximum efficiency was precisely 96 %, and the proposed converter was suitable for the integration of individual photovoltaic cells to 480V.

## ACKNOWLEDGEMENT

The author would gratefully thank Al Jouf University, Sakakah in Saudi Arabia, center of power electronics and department of EECS at University of Missouri-Columbia for their support.